\documentclass[twocolumn]{aastex631}

\usepackage{amsmath}
\usepackage{enumitem}

\usepackage[table]{xcolor}
\usepackage{colortbl}
\usepackage{soul}

\usepackage{booktabs}
\usepackage{multirow}
\usepackage{array}
\usepackage{tabularx}
\usepackage{tabularray}

\shortauthors{Lane, Stephan et al.}

\begin{document}

\title{Observable Metal Pollution in Main-Sequence Stars:\\ Simulations of Rocky Planets Engulfed by Stars in the $0.5$ to $1.4$~M$_\odot$ Range}

\correspondingauthor{Alexander P. Stephan}
\email{alexander.stephan@vanderbilt.edu}

\author[0009-0006-7578-0226]{Kaitlyn T. Lane}
\affiliation{Department of Physics and Astronomy, Vanderbilt University, Nashville, TN 37235, USA}

\author[0000-0001-8220-0548]{Alexander P. Stephan}
\affiliation{Department of Physics and Astronomy, Vanderbilt University, Nashville, TN 37235, USA}

\author[0000-0001-7493-7419]{Melinda Soares-Furtado}
\affiliation{Department of Astronomy,  University of Wisconsin--Madison, 475 N.~Charter St., Madison, WI 53706, USA}
\affiliation{Department of Physics, University of Wisconsin--Madison, 1150 University Ave, Madison, WI 53706, USA}
\affiliation{Wisconsin Center for Origins Research, University of Wisconsin--Madison, 475 N Charter St, Madison, WI 53706, USA}

\author[0000-0002-3481-9052]{Keivan G. Stassun}
\affiliation{Department of Physics and Astronomy, Vanderbilt University, Nashville, TN 37235, USA}

\author[0000-0003-0381-1039]{Ricardo Yarza}
\affiliation{Department of Astronomy and Astrophysics, University of California, Santa Cruz, CA 95064, USA}

\begin{abstract}
The engulfment of planets by their host stars is an expected outcome of various dynamical processes and has been invoked to explain a variety of observed stellar properties, such as rapid rotation, chemical abundance abnormalities, and other transient phenomena. Recent observations support engulfment as the cause of such signatures; however, many engulfment process details remain uncertain. Here, we present a model for determining the chemical signatures produced due to the pollution of main sequence stars by rocky planets, a common engulfment scenario due to the high frequency of observed short-period rocky exoplanets. A key novel element of our model is that we calculate the gradual evaporation of the planet due to drag interactions with the stellar envelope, which can lead to observable pollution on the stellar surface even if the bulk of the planet is only destroyed below the star's outer convective zone. Our results indicate that rocky planet pollution is most easily measurable for stars in the $1.0$ to $1.4$~M$_\odot$ range and that elements such as aluminium, calcium, and vanadium, in addition to lithium, are most suited to detect pollution. We predict that it is also possible to differentiate between the engulfment of one large planet versus several small planets, for the same total pollution mass, for some stellar hosts. We find that rocky planet engulfment events generally take years to decades for most stars. Our results can guide future observational campaigns that may search for sites of past or current engulfment events.
\end{abstract}


\section{Introduction} \label{sec:intro}

Over recent decades, exoplanets have been observed to orbit a diverse set of host stars, spanning all stages of stellar evolution and a large range of stellar masses \citep[e.g.,][]{WolszczanFrail1992,Charpinet+2011,Johnson+2011A,Gettel+2012,Howard+2012,Vanderburg+2020}. An expected outcome of a variety of planetary dynamical processes has been the engulfment and destruction of many of these exoplanets by their host stars \citep[e.g.,][]{Chatterjee+2008,Veras2016,Stephan+2017,Stephan+2018}; indeed, various studies have found evidence for such engulfment events, for example, in the form of stellar spin-up \citep[e.g.,][]{Mamajek+2008,Meibom+2009,James+2010,Barnes+2015,Qureshi+2018, Ong+2024}, white dwarf (WD) metal pollution \citep[e.g.,][]{Zuckerman+2003,Jura2008, Gaensicke+2012,Koester+2014}, chemical composition differences between binary and co-moving stars \citep[e.g.,][]{Oh+2018,Liu+2024}, or lithium enrichment \citep[e.g.,][]{Aguilera+2016,Soares-Furtado+2021}. The most direct evidence for planetary engulfment was discovered recently in the form of an active engulfment event showing significantly increased infrared emission by the star during the event \citep[][]{De+2023,Lau+2025}, consistent with model predictions \citep[e.g.,][]{Metzger+2012,Stephan+2020}.

As more evidence for planetary engulfment as a common astrophysical process has emerged, it has become crucial to develop a thorough understanding of the mechanisms and prevalence of such engulfment events. Some works have begun to make occurrence rate estimates based on observations, suggesting that a significant fraction of stars show chemical signatures from these engulfment events \citep[e.g.,][]{Liu+2024,Spina2024}; however, most studies were so-far only focused on Sun-like stars. Other works have tackled the engulfment process itself from a theoretical and numerical perspective, studying the hydrodynamic processes that occur during the engulfment event \citep[e.g.,][]{Sandquist+1998,JiaSpruit2018,Yarza+2023}; however, many of these works have, in particular, focused on the engulfment of gas giant planets by red giant stars.

In this work, we are developing an analytical approach for estimating and comparing the theoretical limits for observing chemical enrichment signatures in main sequence (MS) stars in the mass range from $0.5$ to $1.4$~M$_\odot$, with the aim of determining which stars are most and least suited for detecting planetary engulfment signatures. We are focusing on the engulfment of rocky, terrestrial planets, which constitute a large fraction of known short-period planets \citep[e.g.,][]{Gaudi+2021}, as determining the detectability of such a signature directly contributes to the estimate of the occurrence rate of potentially Earth-like planets. Additionally, such observations can also measure the chemical composition of terrestrial exoplanets, as has been done before with WD pollution \citep[e.g.,][]{JuraYoung2014,Bonsor2024}, giving key insights into the likelihood that terrestrial exoplanets are indeed Earth-like. We also conduct calculations for stars up to a stellar mass of $2.0$~M$_\odot$; however, their results are not straightforward to interpret, as we show in later sections.

\begin{figure*}
    \centering
    \includegraphics[width=1\linewidth]{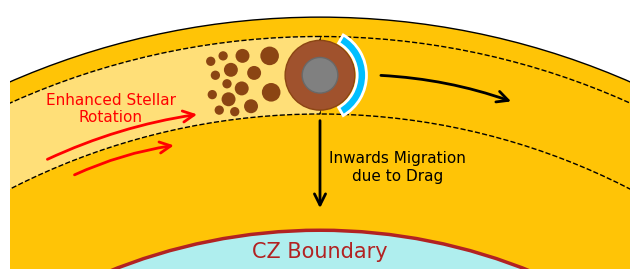}
    \caption{{\bf Schematic of the Engulfment Process.} The figure shows a not-to-scale, zoomed in illustration of an engulfed planet (drawn with a gray center and brown shell to represent the metallic core and rocky mantle) as it travels through the stellar envelope (from left to right, highlighted by the curved black arrow). As the planet travels through the stellar CZ (in yellow) , it is gradually evaporated and slowed down by drag interactions at the shock front (blue curve), sinking deeper into the star over time. The planet heats and spins up the stellar material it interacts with along its path (marked by the dashed curves) and enriches it with the evaporated material (represented by the particles coming off of the planet). Depending on the sizes of the planet and the star, the planet may eventually sink below the CZ boundary into the interior radiative zone of the star, marked by the lower light blue region.}
    \label{fig:evap_graphic}
\end{figure*}

To determine the detectability of engulfed planetary material by stars, we model the engulfment event by calculating the orbital trajectory of the planet in the stellar envelope under the effects of drag, tides, and planetary mass loss via ablation and ram-pressure disintegration (see Fig.~\ref{fig:evap_graphic} for a schematic plot of the scenario). A key element of our approach is that we do not consider 3D-hydrodynamical effects, but we keep our model analytical and one-dimensional; while this approach naturally sacrifices accuracy, one of our goals is to be able to compare engulfment events across a wide variety of stellar hosts within a short simulation time. Importantly, unlike most previous works, we carefully consider the gradual mass loss due to evaporation of the planetary surface, which can add planetary material to a star's outer convective zone (CZ) even if most of the planet is only disrupted after the planet sinks below the CZ boundary into the radiative interior. We determine how much planetary material is added to the stars' outer CZs from the engulfment of a given terrestrial planet and compare it to the primordial chemical composition of the star, using the Sun's composition as our baseline. The basic scenario under consideration here is that planetary pollution can produce elemental abundance differences in stars that ought to have been born with similar initial compositions, such as binary or co-moving star pairs or stellar clusters, as well as composition differences in a single star that may be measurable after an observed engulfment event. The details of our model are explained in Section~\ref{sec:methods}. We present the results of our calculations in Section~\ref{sec:results} and discuss the implications for engulfment detectability and future steps in Section~\ref{sec:discussion}. 

An important insight of our results is that the chemical enrichment of the outer stellar layers by a single Earth-like planet is most straightforwardly detectable for MS stars in the $1.2$ to $1.4$~M$_\odot$ range due to their thin outer CZs, while for mostly convective stars below $\sim0.8$~M$_\odot$, tens to hundreds of Earth masses of planetary material are required for a significant detection, which is also consistent with previous studies of lithium pollution of MS stars via gas giants \citep[e.g.,][]{Soares-Furtado+2021}. Stars above $\sim1.4$~M$_\odot$ have outer CZs of negligible mass, and the detection of planetary material depends primarily on the atomic diffusion rate in their radiative layer; however, the planetary engulfment event itself may significantly alter the stellar structure of such stars due to the added material and extended interaction with the stellar surface. As such, our model can currently make only limited predictions for MS stars with masses above $1.4$~M$_\odot$.

\section{Mathematical Methods} \label{sec:methods}

To accurately and efficiently model the engulfment of the planet by the stellar envelope, we focus on three crucial physical and chemical processes and keep track of them in our simulation: 1.~The orbital evolution of the planet as it undergoes drag and tidal forces; 2.~The gradual evaporation and eventual disruption of the planet via interactions with the stellar envelope material; 3.~The transfer of angular momentum from the planet into the stellar envelope, leading to partial stellar ``spin-up''. Here, we outline our model calculations and the equations used to describe these processes. 

\begin{figure}
    \centering
    \includegraphics[width=1\linewidth]{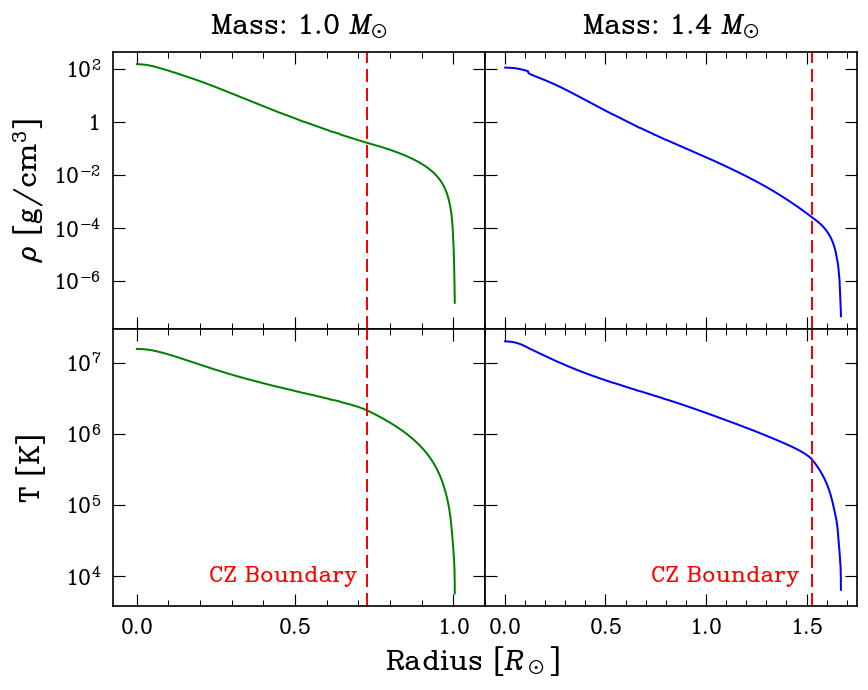}
    \caption{{\bf Radial Stellar Density and Temperature Profile Examples.} We use {\tt MESA}-generated radial stellar structure profiles for our simulations. Here, we show density and temperature profile examples, for main-sequence stars with mass $1$ (left frames, green lines) and $1.4$~M$_\odot$ (right frames, blue lines). The dashed red line marks the inner boundary of the outer convective zone. We use standard parameter inputs for the calculations (see online {\tt MESA} documentation).}
    \label{fig:profiles}
\end{figure}

To have realistic stellar parameters for our calculations, we used the {\tt MESA} code \citep[Modules for Experiments in Stellar Astrophysics,][, release version 12778]{Paxton+2011, Paxton+2013, Paxton+2015, Paxton+2018, Paxton+2019, Jermyn+2023}, which provides stellar density and temperature profiles, as well as an estimate for the boundary between the convective and radiative layers of a given star, via the Schwarzschild criterion (see example profiles in Fig.~\ref{fig:profiles}). We focus on stars in the $0.5$ to $2$~M$_\odot$ mass range with initially solar metallicity \citep[$Z=0.02$, see also][]{AndersGrevesse1989, Asplund+2021}. We use {\tt MESA} profiles that correspond to the age at which the stars are securely on the MS (i.e. they are no longer shrinking) for the $0.5$ to $0.8$~M$_\odot$ stars and profiles that are roughly at the mid-point of the stars' MS lifetimes for $0.9$ to $2$~M$_\odot$ stars for our main calculations. We use profiles corresponding to the early and late MS stages to estimate a range of feasible values that may be observed over the course of a star's MS lifetime (see Table~\ref{tab:stellar_properties} for the ages of the chosen profiles).


Regarding the planet, we primarily focus on the scenario of a rocky planet of similar size, mass, and composition to Earth, but we also test our model with a Super Earth with mass $15$~M$_{\oplus}$ and radius $\approx 2.47$~R$_{\oplus}$. For simplicity, we assume the planet to have uniform density, but we do consider the crust, mantle, and core fractions of the planet for its composition.

\subsection{Orbital Evolution of the Planet} \label{subsec:orbit}

The orbital evolution of the planet during the engulfment event is determined by two main forces: drag from the stellar envelope and tidal interactions with the star at large. We are agnostic about the process that brought the planet into the envelope; however, our calculations mostly reflect the case for a low-eccentricity, ``gentle'' process, with tidal decay bringing the planet to the stellar surface initially.

Drag will reduce the angular momentum $L$ of the planet's orbit via
\begin{equation}
    \label{eq:drag}
    \frac{dL}{dt}\Bigr|_{\text{drag}} = \dot{L}_{\text{drag}} =
    -\frac{1}{2}\, c_{\text{d}}\, a_{\text{p}}\, \rho_{\ast}\, r_{\text{in}}\, v_{\text{rel}}^2,
\end{equation}
where $c_{\text{d}}$ is a dimensionless drag coefficient of value $0.5$, $\rho_{\ast}$ is the density of the stellar material at the distance of the planet from the star's center, $r_{\text{in}}$, $v_{\text{rel}}$ is the relative speed between the planet and the surrounding stellar material ($v_{\text{rel}}=v_{\text{orb}}-v_{\text{spin}}$), and  
$a_{\text{p}}$ is the effective cross-section of the planet, which is simply $\pi\,r_{\text{p}}^2$ when the scale height, $h$, is larger than the planet's radius, $r_\text{p}$, but is $\pi\,r_{\text{p}}^{1/2}\,h^{3/2}$ otherwise \citep[see also][]{Metzger+2012,Stephan+2020}. The scale height $h$ of the stellar gas determines the length scale across which the stellar gas density increases significantly and is calculated as 
\begin{equation}
    \label{scale_height}
    h = \frac{k_{B}\, T\, r_{\text{in}}^2}{\mu\, m_{H}\, G\, m_{\text{in}}},
\end{equation}
where $k_B$ is the Boltzmann constant, $T$ is the temperature of the stellar envelope at the position of the planet, $\mu\,m_H$ is the average atomic mass of the stellar gas particles, $m_{in}$ is the mass of the star interior to the planet's position, and $G$ is the gravitational constant. 

We treat tidal effects with a simplified approach following the equilibrium tide model \citep{Hut1981,1998EKH}, where tidal forces will reduce the planet's angular momentum via 
\begin{equation}
    \label{eq:tide}
    \frac{dL}{dt}\Bigr|_{\text{tides}} = \dot{L}_{\text{tides}} = -\frac{L}{2 t_{\text{f}}} \left(1-\frac{\Omega}{\omega}\right),
\end{equation}
with $\Omega$ being the stellar spin rate, $\omega$ being the planet's orbital rate, and $t_{\text{f}}$ being the tidal friction timescale 
\begin{equation}
    \label{eq:timescale}
    t_{\text{f}} = \frac{t_{\text{v}}}{9} \left(\frac{a_{\text{orb}}}{r_{\text{in}}}\right)^8 \frac{m_{\text{in}}^2}{(m_{\text{in}}+m_{\text{p}}) m_{\text{p}}} \frac{1}{(1+k_{\text{L}})^2},
\end{equation}
where t$_{\text{v}}$ is the viscous timescale (on the order of years for stars), $k_{\text{L}}$ is the Love parameter, $a_{\text{orb}}$ is the planet's orbital semi-major axis (approximately equal to $r_{\text{in}}$, which thus cancel here), and $m_{\text{p}}$ is the planet's mass. Initially, we set the stellar spin period to $20$~days for $0.5 - 1.4$~M$_{\odot}$ stars and $2$~days for $1.5 - 2$~M$_{\odot}$ stars to reflect the effects of the Kraft break \citep{Kraft1967} and the expected spin-down of stars with deep CZs \citep[see, for example][the Sun has a current spin period of about 25 days]{Barnes2003,Barnes+2015}. The spin rate of the stellar envelope changes via interactions with the planet, which we expand upon further in Sec.~\ref{subsec:impact}.

{We note here that our treatment of tidal effects is an approximation. As the planet migrates inwards, one should expect the tidal response to deviate from the equilibrium tide model. However, tides are most important when the planet is in the outermost layers of the star, where tidal dissipation generally dominates over dissipation by drag. In these outer layers, our choice of tidal model should provide adequate results, while interior to them, tidal effects quickly become negligible in our calculations.}

From the angular momentum changes due to drag and tides, we can derive the (secular) inwards migration speed of the planet. Given that
\begin{equation}
    \label{eq:r_in}
     r_{\text{in}} = \frac{L}{m_{\text{p}} v_{\text{orb}}},
\end{equation}
we find that the inward radial speed is
\begin{equation}
    \label{eq:drdt}
    v_{\text{r}} = \frac{dr_{\text{in}}}{dt} = \frac{d}{dt}\left(\frac{L}{m_{\text{p}} v_{\text{orb}}}\right) = \frac{\dot{L}}{m_{\text{p}} v_{\text{orb}}}+\frac{L}{m_{\text{p}}}\frac{d}{dt}\frac{1}{v_{\text{orb}}},
\end{equation}
where $v_{\text{orb}}$ is the orbital speed of the planet. Further, using our previous equations, we obtain the expression
\begin{equation}
    \label{eq:v_r}
    v_{\text{r}} = \frac{  \left(\dot{L}_{\text{drag}}+\dot{L}_{\text{tides}}\right)/\left(m_{\text{p}} v_{\text{orb}}\right) 
    }{   1 + \frac{G}{2 v_{\text{orb}}^2} \left( \dfrac{dm_{in}}{dr}\Bigr|_{r_{in}} - \dfrac{m_{\text{in}}}{r_{\text{in}}}\right) },
\end{equation}

where $dm_{in}/dr$ is the radial change of the stellar mass that is interior to the planet's orbit as the planet migrates inwards, determined at $r_{in}$. As the density of the stellar material surrounding the planet rapidly increases as the planet gets deeper, the inwards migration due to drag eventually becomes a (non-secular) ``plunge'', which we show in later sections. 

\setlength{\tabcolsep}{2.25pt}
\begin{table*}
\parbox[t][][t]{0.36\textwidth}{
\caption{{\bf Stellar Ages of utilized {\tt MESA} Profiles.} The table lists the ages of the stellar {\tt MESA} profiles we used for the investigated stellar masses. The early and late MS profiles were used to estimate the range of possible values in our model calculations, as seen in Fig. \ref{fig:summary_plot}.}
\label{tab:stellar_properties}
\centering
\begin{tabular}{|c|c|c|c|}
\hline
\vtop{\hbox{\strut Mass} \hbox{\strut (M$_{\odot}$)}} &
\vtop{\hbox{\strut Early MS} \hbox{\strut Age (Gyr)}} &
\vtop{\hbox{\strut \textbf{Main MS}} \hbox{\strut \textbf{Age (Gyr)}}} & 
\vtop{\hbox{\strut Late MS} \hbox{\strut Age (Gyr)}} \\
\hline
0.5 & 0.163 & \textbf{4.300} & 57.00 \\
0.6 & 0.122 & \textbf{0.404} & 36.00 \\
0.7 & 0.090 & \textbf{0.583} & 24.00 \\
0.8 & 0.068 & \textbf{0.256} & 17.00 \\
0.9 & 0.132 & \textbf{7.458} & 13.00 \\
1.0 & 0.043 & \textbf{4.790} & 9.878 \\
1.1 & 0.050 & \textbf{3.145} & 3.402 \\
1.2 & 0.032 & \textbf{2.798} & 4.584 \\
1.3 & 0.024 & \textbf{2.215} & 2.360 \\
1.4 & 0.019 & \textbf{1.717} & 2.719 \\
1.5 & 0.067 & \textbf{1.000} & 1.419 \\
1.6 & 0.041 & \textbf{1.093} & 1.189 \\
1.7 & 0.035 & \textbf{0.884} & 1.514 \\
1.8 & 0.027 & \textbf{0.758} & 1.280 \\
1.9 & 0.022 & \textbf{0.645} & 1.103 \\
2.0 & 0.019 & \textbf{0.525} & 0.962 \\
\hline
\end{tabular}
}
\hfill
\parbox[t][][t]{0.6\textwidth}{
\caption{{\bf Abundances of Key Elements in the Sun and Earth.} The table lists the abundances by percentage of a selection of key elements found in rocky planets, for the Sun \citep{Asplund+2021} and for Earth. The abundances for Earth are further differentiated between the crust \citep{Haynes2016}, mantle \citep{Palme2014}, and core \citep{McDonough2003}, with the caveat that the core composition is based on model estimates.}
\label{tab:abundances}
\centering
\begin{tabular}{|c|l|l|l|l|l|}
\hline
Element & Sun & \vtop{\hbox{\strut Earth} \hbox{\strut Crust}} & 
\vtop{\hbox{\strut Earth} \hbox{\strut Mantle}} & \vtop{\hbox{\strut Earth} \hbox{\strut Core}} & \vtop{\hbox{\strut Bulk} \hbox{\strut Earth}} \\
\hline
Li & $4.67\times10^{-9}$ & $2.00\times10^{-3}$ & $1.60\times10^{-4}$ & - & $1.29\times10^{-4}$ \\
O & 0.58 & 46.10 & 44.33 & - & 30.61 \\
Na & $2.82\times10^{-3}$ & 2.36 & 0.26 & - & 0.20\\
Mg & 0.06 & 2.33 & 22.17 & - & 15.10 \\
Al & $5.36\times10^{-3}$ & 8.23 & 2.38 & - & 1.70 \\
Si & 0.07 & 28.20 & 21.22 & 6.00 & 16.57 \\
S & 0.03 & 0.04 & 0.02 & 1.90 & 0.60\\
K & $3.39\times10^{-4}$ & 2.09 & 0.03 & - & 0.04 \\
Ca & $5.90\times10^{-3}$ & 4.15 & 2.61 & - & 1.82 \\
Ti & $3.29\times10^{-4}$ & 0.57 & 0.13 & - & 0.09 \\
V & $2.99\times10^{-5}$ & 0.01 & $8.60\times10^{-3}$ & 0.02 & 0.01 \\
Cr & $1.60\times 10^{-3}$ & 0.01 & 0.25 & 0.90 & 0.45 \\
Mn & $1.07\times 10^{-3}$ & 0.10 & 0.11 & 0.03 & 0.09 \\
Fe & 0.12 & 5.63 & 6.30 & 85.50 & 30.84 \\
Ni & $6.86\times10^{-3}$ & $8.40\times10^{-3}$ & 0.17 & 5.20 & 1.73 \\
Y & $1.06\times 10^{-6}$ & $3.30\times 10^{-3}$ & $4.13\times 10^{-4}$ & - & $3.14\times10^{-4}$ \\
\hline
\hline
\vtop{\hbox{\strut Total Body} \hbox{\strut Mass (g)}} & $1.989\times 10^{33}$ & $5.97\times 10^{25}$ & $4.06\times 10^{27}$ & $1.85\times 10^{27}$ & $5.97\times 10^{27}$ \\
\hline
\end{tabular}
}
\end{table*}

\subsection{Evaporation and Disintegration of the Planet} \label{subsec:evaporation}

As the planet migrates deeper into the stellar envelope, the drag it experiences will gradually remove planetary material via evaporation and eventually disrupt the entire planet. For our model, we calculate the energy that is dissipated at the envelope-planet shock-front interface and determine how much mass it can unbind from the planetary surface, considering both the gravitational and chemical binding energy of the material.

The planet moves through the stellar envelope typically well above the local sound speed, $c_\text{s}$, as determined by our {\tt MESA} models, which implies that the envelope-planet interface is a shock front with a high temperature in the tens of thousands of Kelvin \citep[e.g.,][]{Metzger+2012}. As such, we assume that the energy dissipated in the shock front can efficiently evaporate the planetary material. As a simple model, we compare the mass change per unit energy, $dm_{\text{p}}/dE$, with the available energy over time from the drag interactions, $dE/dt|_{\text{drag}}=dL/dt|_{\text{drag}}\,v_\text{rel}$. Naturally, only a fraction of the thermal energy in the shock front should be available to evaporate the planetary surface, which we estimate by approximating the interactions between the stellar envelope particles and the planetary material particles as elastic collisions. This results in the energy being split between them according to the partition function
\begin{equation}
    \label{eq:Partititon}
    F \approx \frac{\mu}{\mu+60u},
\end{equation}
where $\mu$ is the average atomic mass of the stellar material ($\sim0.62u$ for a full plasma). We assume that the planetary material has an average atomic mass of $~60u$, since a rocky planet like Earth consist predominantly of iron (Fe, atomic weight $55.845u$) and silicon dioxide (SiO$_2$, molecular weight $60.08u$). As such, we expect that only a small fraction of the energy, approximately $1$~\%, goes into evaporating the planet, ignoring radiative energy transfer.

The energy available from the drag interaction will have to overcome the gravitational and chemical binding energy of the planetary material. We define these via 
\begin{equation}
    \label{eq:E_bind}
    \varepsilon_{\text{grav}} \approx \frac{G m_\text{p}}{R_\text{p}}, 
\end{equation}
the gravitational binding energy density per unit mass of planetary material, and
\begin{equation}
    \label{eq:E_evap}
    \varepsilon_{\text{chem}} \approx 10^7 \text{Joule}/\text{kg},
\end{equation}
the chemical binding energy density per unit mass of planetary material. The value chosen for the chemical binding energy density is a conservative estimate based on the experimentally determined evaporation energy of various kinds of rocky material \citep{Navrotsky1995}; this value is generally much lower than the value of the gravitational binding energy density, but it is non-negligible.

Combining these previous equations, we arrive at the expression for the planetary mass lost via the drag,
\begin{equation}
    \label{eq:m_loss_evap}
    \frac{dm_{\text{p}}}{dt}\Bigr|_{\text{drag}} = \frac{dE}{dt}\Bigr|_{\text{drag}}\frac{dm_{\text{p}}}{dE} = \frac{1}{2}
    \frac{ c_{d}\, a_{\text{p}}\, \rho_{\ast}\, |v_{\text{rel}}^3|\,  F}{\varepsilon_{\text{grav}} + \varepsilon_{\text{chem}}}.
\end{equation}
This mass loss equation holds as long as the conditions in the stellar envelope do not lead us to expect that the planet would break apart completely. In general, we expect the planet to start disintegrating when the ram pressure of the drag interface overcomes the remaining gravitational binding energy of the whole planet; this condition can be expressed in terms of the stellar and planetary densities, $\rho_{\ast}$ and $\rho_{\text{p}}$, respectively, the planet's speed through the stellar envelope, $v_{\text{rel}}$, and the planet's escape velocity, $v_{\text{p,esc}}$, via
\begin{equation}
    \label{eq:dis}
    \rho_{\ast} > \rho_{\text{p}} \left( \frac{v_{\text{p,esc}}}{v_{\text{rel}}} \right)^2,
\end{equation}
\citep[see also][]{JiaSpruit2018}. Once this disruption condition is fulfilled, the planet will start to rapidly disintegrate, losing its remaining mass on a typical timescale of 
\begin{equation}
    \label{eq:t_dis}
    t_{\text{dis}} = \frac{1}{\sqrt{G\rho_{\text{p}}}},
\end{equation}
leading to a mass loss equation of
\begin{equation}
    \label{eq:m_loss_dis}
    \frac{dm_{\text{p}}}{dt}\Bigr|_{\text{dis}} =
    - \frac{m_{\text{p}}}{t_{\text{dis}}}.
\end{equation}
We note here that the Roche limit, at which the planets would be ripped apart by tidal forces from the star, is deep within the stellar interior for the rocky planets under consideration, and in all our scenarios, the planets undergo complete evaporation and disintegration before they can reach the Roche limit.

\subsection{Impact of the Planetary Material on the Envelope} \label{subsec:impact}

As we have highlighted in the previous sections, the relative velocity of the planet compared to the envelope material is important for the calculations of the drag effects and not just the orbital velocity. As such, the inherent spin velocity of the stellar material needs to be estimated, and changes in the stellar spin due to the transfer of angular momentum from the planet's orbit to the envelope need to be considered.

We initially assume that the stars rotate at a uniform rate of $20$~days for $0.5 - 1.4$~M$_{\odot}$ stars and $2$~days for $1.5 - 2$~M$_{\odot}$ stars; stars below the Kraft break are expected to spin down significantly over their MS lifetimes, while stars above the break remain spinning at short periods for most of their lives \citep[e.g.,][]{Kraft1967,Barnes2003}, with the Sun's current spin period being $25$~days. A conservative way to include the effects of stellar spin-up is to keep treating the star as a uniformly rotating body and add the transferred angular momentum to it, such that the change in the spin rate can be calculated as
\begin{equation}
    \label{new_omega_con}
    \frac{d\omega_{\text{uniform}}}{dt} = \left(\frac{dL}{dt}\Bigr|_{\text{drag}}+\frac{dL}{dt}\Bigr|_{\text{tides}}\right)/I,
\end{equation}
assuming a constant moment of inertia $I$. However, in general, stars cannot be expected to rotate as solid bodies, with observations and theory indicating that stars show both latitudinal and radial variations in rotation rate \citep[e.g.,][]{GoldreichSchubert1967,Duvall+1984,Zahn1992,Schou+1998}. As such, we are estimating the spin-up of the stellar material by approximating the path of the planet as a torus, to which we add the angular momentum transferred via drag to determine the increased spin velocity.

For the purposes of calculating the new spin velocity of the material accelerated via drag, we estimate the diameter of the (smaller) torus circle based on the size of the planet and the distance the planetary material that interacted with the drag can be expected to diffuse over the course of a single orbital period, $r_{\text{eff}} = r_{\text{p}} + d_{\text{diff}}$, as the planet will approximately return to this material again after that time. As such, we use the mean free path, $l$, and the thermal (average) velocity, $v_{\text{th}}$, for the diffusing particles,
\begin{equation}
    \label{mfp}
    l = \frac{k_{B}T}{\sqrt{2\,} \pi {(d_{\text{ptcl}})}^2\,P},
\end{equation}
and
\begin{equation}
    \label{v_thermal}
    v_{\text{th}} = \sqrt{\frac{2\, k_{B}T}{m_{\text{ptcl}}}},
\end{equation}
where $d_{\text{ptcl}}$ is the effective diameter of the diffusing particles, $m_{\text{ptcl}}$ is the mass of the diffusing particles, which we assume to be iron for this calculation, and $P$ is the stellar gas pressure. Using $l$ and $v_{\text{th}}$, we determine the average time between particle collisions,
\begin{equation}
    \label{t_coll}
    t_{\text{coll}} =\frac{l}{v_{\text{th}}} = \frac{\sqrt{k_{B}\, T\, m_{\text{ptcl}\,}}}{2\, \pi {(d_{\text{ptcl}})^2\, P}},
\end{equation}
which in turn gives us the diffusion length scale over a single orbital period,
\begin{equation}
    \label{d_diff}
    d_{\text{diff}}(t_{\text{orb}}) = l \sqrt{\frac{t_{\text{orb}}}{t_{\text{coll}}}} = {\left( \frac{{(k_{B}T)}^3}{m_{\text{ptcl}}} \right)}^{\frac{1}{4}} \frac{\sqrt{t_{\text{orb}}}}{{d_{\text{ptcl}}\sqrt{\pi P\,}}} .
\end{equation}
We add the angular momentum lost by the planet via drag to the torus determined by $r_{\text{eff}} = r_{\text{p}} + d_{\text{diff}}$ and use the moment of inertia for a torus to calculate the \textit{added} spin velocity. This added spin velocity, together with the inherent spin velocity of the star, is used to then calculate the relative velocity of the planet through the stellar material, as outlined in previous sections.

However, as the planet keeps traveling through the star's outer CZ, it will keep sinking deeper into the star and will not actually encounter exactly the same material again after each orbit. Furthermore, the accelerated material and its angular momentum will keep diffusing over many orbital periods. These two factors put limits on the spin velocity of the material the planet will encounter. We determine an approximate diffusion velocity by taking the derivative of the diffusion length with respect to the orbital period, such that
\begin{equation}
    \label{v_diff}
    v_{\text{diff}}(t_{\text{orb}}) = {\left( \frac{{(k_{B}T)}^3}{m_{\text{ptcl}}} \right)}^{\frac{1}{4}} \frac{1}{2\,d_{\text{ptcl}}\,\sqrt{\pi\, P\,t_{\text{orb}}}}.
\end{equation}
With this diffusion velocity $v_{\text{diff}}$, we can then approximate the rate at which angular momentum leaves the effective torus of the planet's path, 
\begin{equation}
    \label{dldt_diff}
    \frac{dL}{dt}\Bigr|_{\text{diff}} = - \frac{2\, L\, v_{\text{diff}}}{r_{\text{p}}} \left(1 + \frac{d_{\text{diff}}}{r_{\text{p}}} \right),
\end{equation}
and further reduce the angular momentum due to the planet's sinking via 
\begin{equation}
    \label{dldt_sink}
    \frac{dL}{dt}\Bigr|_{\text{sink}} = \frac{L\, v_{\text{r}}}{2\, r_{\text{eff}}}.
\end{equation}

We note that if the planet sinks faster than its orbital speed, $v_{\text{orb}}<v_{\text{r}}$, such as during the final ``plunge'', then the planet will not encounter any of the material it has previously accelerated, and only the initial spin of the stellar envelope is relevant for our calculations.

{Finally, we point out here that our above treatment of the diffusion processes is simply an approximation of the much more complicated underlying fluid and plasma mechanics. We ignore the effects of turbulent diffusion, because even though the convection speed is very high within the outer convective zone, we estimate the convective cells themselves to be smaller than the size of the engulfed planets, thus not contributing much to actually carrying away evaporated material from the planet's migration path. } 

\begin{figure*}
    \centering
    \includegraphics[width=1\linewidth]{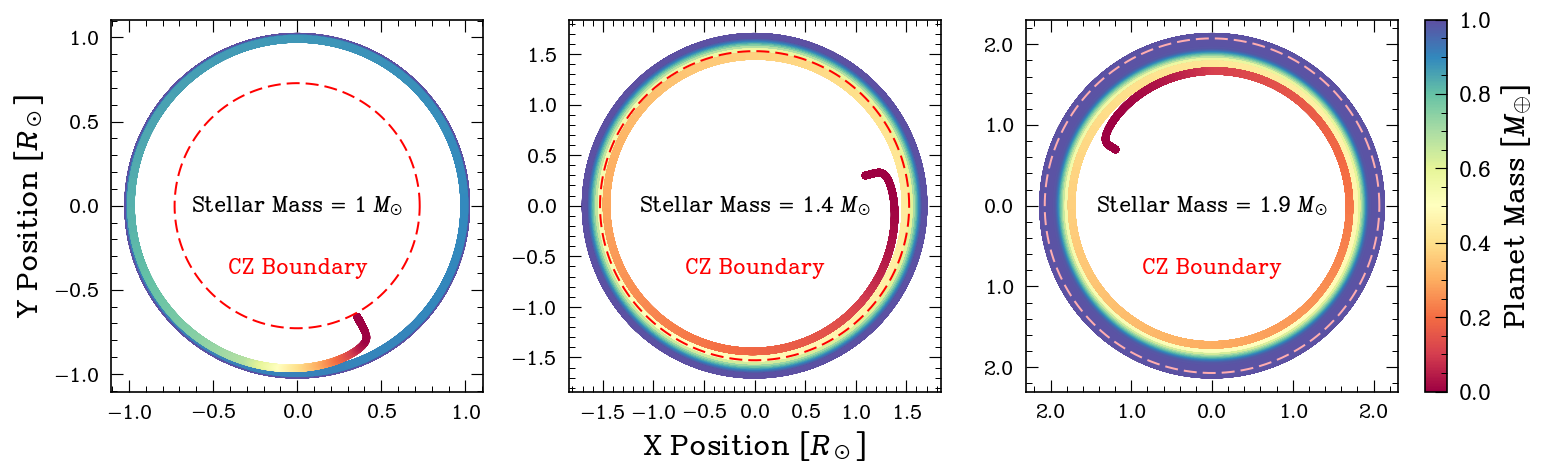}
    \caption{{\bf In-spiral Trajectory of Engulfed Planets.} The three figures show the orbital paths an Earth-like planet travels during its engulfment by a MS star. The left, center, and right plots show examples for $1$, $1.4$, and $1.9$~M$_\odot$ stars, respectively. The colors highlight the remaining mass of the planet at a given position on this trajectory, with blue indicating an intact planet and red a nearly completely destroyed one. The inner boundary of the outer CZ is marked by the red dashed circles. Note that the planet is completely destroyed within the CZ for the $1$~M$_\odot$ case, but reaches deeper into the stellar interior for the $1.4$ and $1.9$~M$_\odot$ cases.}
    \label{fig:orbit1}
\end{figure*}

\begin{figure*}
    \centering
    \includegraphics[width=1\linewidth]{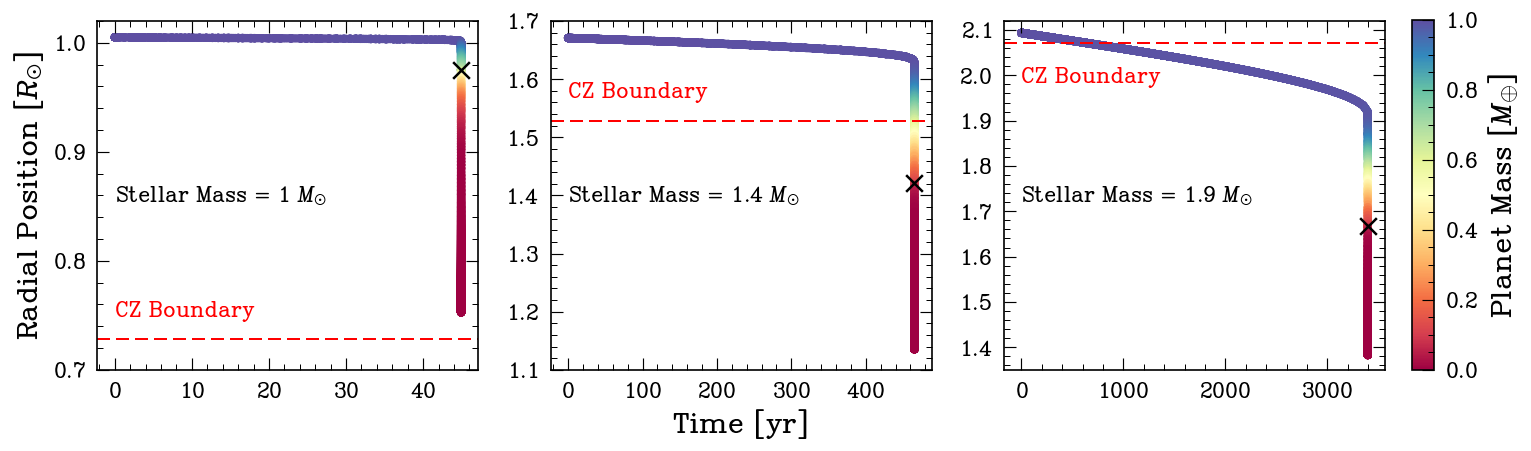}
    \caption{{\bf Orbital Distance Evolution of Engulfed Planets.} The three figures show the evolution of the orbital distance of an Earth-like planet during its engulfment by a MS star. The left, center, and right plots show examples for $1$, $1.4$, and $1.9$~M$_\odot$ stars, respectively. The colors have the same meaning as in Fig.~\ref{fig:orbit1}. The inner boundary of the outer CZ is marked by the red dashed lines. The black markers highlight the position at which a planet's disruption condition is reached; before this point, the planet can still be considered to be a consolidated object, while after this point, the planet quickly disintegrates into a rubble pile and evaporates.}
    \label{fig:plunge1}
\end{figure*}

\subsection{Composition of the Star and Planet}\label{subsec:planetcomp}

As one of our goals is to determine whether planetary material engulfed by a given star can produce detectable changes in the stellar surface composition, we estimate the changes in elemental composition from the amount of added planetary material to the outer CZ. We assume that the stars have an initial composition similar to solar \citep[using data from][]{Asplund+2021}, and we use measured and estimated elemental compositions for the Earth's crust \citep{Haynes2016}, mantle \citep{Palme2014}, and core \citep{McDonough2003}. Table~\ref{tab:abundances} lists the percentage abundances of a list of key elements relevant for rocky planets like Earth. Depending on the fraction of the planet's material that is added to the CZ and the stellar mass of that zone, we calculate the change of the stellar surface composition. We assume that any added material is evenly mixed into the outer CZ over short timescales and that atomic diffusion and settling is insignificant for stars with sufficiently thick envelopes.

\section{Results} \label{sec:results}
    
Our simulations highlight three distinct model outcomes: 1.~The planet is fully destroyed in the star's outer CZ; 2.~The planet loses a significant portion of its mass in the star's outer CZ, but reaches the radiative zone below; 3.~The planet loses an insignificant portion of its mass in the star's thin outer CZ, but this amount of material is potentially enough to overwhelm the CZ of the star. We explore these three scenarios using the representative examples of the $1$, $1.4$, and $1.9$~M$_\odot$ star cases, respectively.

\setlength{\tabcolsep}{5pt}
\begin{table*}[!htp]\centering
\caption{{\bf Key Radii and Durations during Engulfment Events.} For the stars considered in our model, the table lists their masses and radii ($M_*$ and $R_*$, respectively), the inner radius of the outer CZ ($R_{CZ}$), the distance from the stellar center at which an engulfed Earth-like (middle columns) or Super-Earth (right columns) planet begins to disrupt ($R_{dis}$), the time at which the planet crosses the CZ, as well as the total duration of the engulfment event. Note that for stars smaller than $1.4$~M$_{\odot}$, an Earth-like planet will begin to disrupt and lose virtually its entire mass before being able to leave the CZ  (below $1.1$~M$_{\odot}$ for Super-Earths).}
\label{tab:engulf_times}
\hspace*{-2.5cm}
\begin{tabular}{|c|c|c| |c|c|c| |c|c|c|}
\hline
\multicolumn{3}{|c||}{} & \multicolumn{3}{c||}{1~M$_{\oplus}$ Planet} & \multicolumn{3}{c|}{15~M$_{\oplus}$ Planet} \\
\hline
$M_*$ (M$_{\odot}$) & $R_*$ (R$_{\odot}$) & $R_{CZ}$ (R$_{\odot}$) &
$R_{dis}$ (R$_{\odot}$) & 
\vtop{\hbox{\strut CZ crossing} \hbox{\strut time (yrs)}} & 
\vtop{\hbox{\strut Engulfment} \hbox{\strut duration (yrs)}} & 
$R_{dis}$ (R$_{\odot}$) & 
\vtop{\hbox{\strut CZ crossing} \hbox{\strut time (yrs)}} & 
\vtop{\hbox{\strut Engulfment} \hbox{\strut duration (yrs)}} \\
\hline
0.5 & 0.452 & 0.265 & 0.450 & - & 1.569 & 0.447 & - & 1.062 \\
0.6 & 0.552 & 0.360 & 0.548 & - & 4.559 & 0.542 & - & 1.636 \\
0.7 & 0.653 & 0.440 & 0.646 & - & 11.75 & 0.635 & - & 2.433 \\
0.8 & 0.721 & 0.499 & 0.712 & - & 17.17 & 0.697 & - & 3.040 \\
0.9 & 0.888 & 0.616 & 0.869 & - & 31.66 & 0.835 & - & 4.674 \\
1.0 & 1.000 & 0.729 & 0.975 & - & 44.92 & 0.920 & - & 6.257 \\
1.1 & 1.140 & 0.886 & 1.089 & - & 67.08 & 0.992 & 9.030 & 9.030 \\
1.2 & 1.335 & 1.109 & 1.238 & - & 111.0 & 1.060 & 16.10 & 16.10\\
1.3 & 1.511 & 1.318 & 1.346 & - & 194.2 & 1.118 & 35.36 & 35.36 \\
1.4 & 1.671 & 1.528 & 1.422 & 465.4 & 465.4 & 1.165 & 109.8 & 109.9 \\
1.5 & 1.681 & 1.645 & 1.362 & 920.3 & 2105 & 1.135 & 66.74 & 297.2 \\
1.6 & 1.905 & 1.885 & 1.510 & 512.2 & 2742 & 1.240 & 35.08 & 364.6 \\
1.7 & 1.953 & 1.934 & 1.552 & 512.5 & 3016 & 1.277 & 34.87 & 392.3 \\
1.8 & 2.032 & 2.011 & 1.615 & 566.0 & 3221 & 1.327 & 38.50 & 418.6 \\
1.9 & 2.094 & 2.072 & 1.667 & 621.9 & 3395 & 1.371 & 42.34 & 441.9 \\
2.0 & 2.112 & 2.090 & 1.692 & 649.3 & 3505 & 1.397 & 44.24 & 456.7 \\
\hline
\end{tabular}
\end{table*}

\subsection{Orbital Evolution}

There are two distinct phases of a given engulfed planet's orbital evolution within a star. In the first phase, the planet orbits inside the star just beneath the stellar surface and retains most of its mass due to the weak drag the planet experiences. This phase is maintained for nearly the entire engulfment process, which can last from the order of years to thousands of years, and is majorly extended due to the reduction in stellar drag from the partial spin-up of the stellar envelope. The second phase begins when the planet starts to migrate deeper into the star and experiences a significant increase in mass loss due to stronger drag. This second phase lasts on the order of the orbital timescale of the planet at its current depth, i.e. hours. Finally, the disruption condition is met, and the planet is fully destroyed in a final plunge into the stellar interior on a timescale of $t_{dis}\approx 0.46$~hours, as estimated from Eq.~\ref{eq:t_dis}. The total engulfment process durations range from about $1.5$ years for a $0.5$~M$_\odot$ star to $3600$ years for a $2.0$~M$_\odot$ star (for a $15$~M$_\oplus$ Super Earth, these durations are reduced to $1$ and $450$~years, respectively; see also Table~\ref{tab:engulf_times}). In Figure \ref{fig:orbit1}, we show the orbital in-spiral trajectories of three examples of the aforementioned scenarios, represented by the $1$, $1.4$, and $1.9$~M$_\odot$ star example cases; these figures highlight that the planet retains most of its mass while near the stellar surface. In Figure \ref{fig:plunge1}, we show the orbital distance evolution over time of our three example cases; these figures highlight the extended period of time the planet remains near the surface versus the short duration of the final plunge and destruction. The innermost radius that the planet reaches by the time of disruption is shown by the black marker; the innermost radius that the disintegrating planetary material may reach is shown by the end of the colored data points.

\subsection{Mass Loss Evolution}

The orbital evolution mentioned previously correlates similarly with the mass loss evolution of the engulfed planets. In particular, during the extended early phase where the planet orbits near the stellar surface, the planet loses very little mass. Most of its mass is lost during the final plunge into the stellar interior. As shown in Figure \ref{fig:plunge1}, a $0.5$ to $1.3$~M$_{\odot}$ star completely disintegrates an Earth-like planet within its CZ, while for stars around $1.4$~M$_{\odot}$, a significant portion of the planet reaches below the CZ. In contrast, for a $1.5$ to $2.0$~M$_{\odot}$ star, most of the planet reaches below the CZ before the disruption condition is met. The division between these scenarios, namely whether or not the planet is fully destroyed in the outer CZ, roughly correlates with the Kraft break at $\sim1.4$~M$_{\odot}$ \citep{Kraft1967}. This is to be expected, as stars above the Kraft break mostly have radiative interiors with very thin, if any, outer CZs. 

\begin{figure*}
    \centering
    \includegraphics[width=1\linewidth]{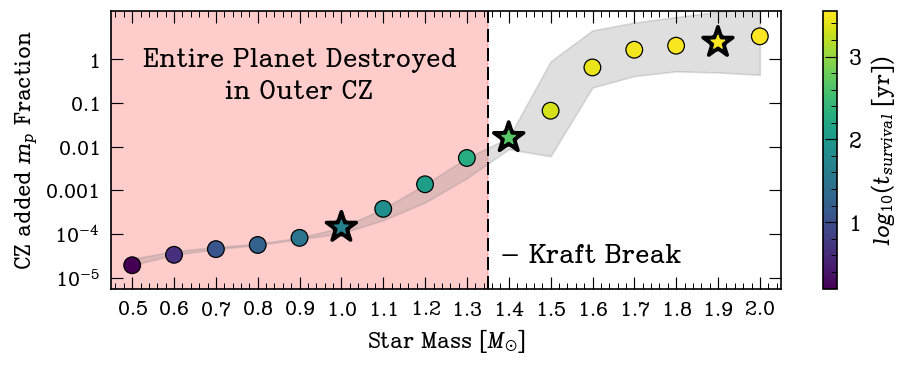}
    \caption{{\bf Comparison between the added Planetary Mass and the Stellar Material Mass in the Outer CZ.} The plot shows how the mass that an Earth-like planet loses to the outer CZ during an engulfment event compares with the initial stellar material mass of that zone. Stars below the Kraft break (approximately marked with the vertical black dashed line) completely destroy an Earth-like planet within their CZs, while for higher mass stars, the planet survives deeper into the stellar interior (see also Table~\ref{tab:engulf_times} and compare to Super-Earth-type planets). The size of the outer convective envelope is also very small for these more massive stars, as they are mostly radiative. As such, while the planet adds a lot of mass (relative to the mass of the CZ) to the CZ for these massive stars, it is uncertain how such a large fraction of added material will impact the structural evolution of the CZ, and we cannot make strong predictions for the observability. It is possible that such significant surface pollution could drive a wind and the production of a dusty shroud around the star \citep[see also][]{Lau+2025}. The dot colors correspond to the total engulfment survival time of the planet. We estimate a range for the added planetary mass fraction by considering the size of the CZ at the beginning versus the end of each star's MS evolution, marked by the gray filled area. The star markers highlight our representative cases of $1$, $1.4$, and $1.9$~M$_\odot$.}
    \label{fig:summary_plot}
\end{figure*}

We compare the planetary mass lost in the CZ to the total stellar mass in the CZ and show the ratio between these masses in Fig.~\ref{fig:summary_plot} (see also Table \ref{tab:mass_fractions} for a more detailed overview). We estimate the range of the ratios that are possible across the MS lifetimes for these stars by running our model with early and late MS {\tt MESA} profiles (shown by the gray shaded area in Fig.~\ref{fig:summary_plot}). 

The ages of the {\tt MESA} profiles we used are listed in Table \ref{tab:stellar_properties}. The $0.5$ to $1.0$~M$_{\odot}$ stars' internal structures and total masses do not change significantly throughout their MS lifetimes, so the fraction of mass the planet loses in the CZ compared to the mass of the CZ itself does \textit{not} have a strong dependence on the star's age. For $1.1$ to $1.4$~M$_{\odot}$ stars, the CZ is larger during the early and late MS phases, yielding a smaller pollution fraction than during the middle of the MS phase (labeled as ``Main MS'' in Table \ref{tab:stellar_properties}). For the $1.5$ to $2.0$~M$_{\odot}$ stars, the early MS profiles yielded the highest pollution fraction, while the late MS profiles yielded the smallest pollution fraction; this is due to the significant enlargement of the CZs as the stars evolve through the MS. However, the CZs are always very small during the MS lifetime, as these stars are above the Kraft break and have nearly completely radiative envelopes. As a result, the pollution fraction is high, potentially overwhelming the stellar mass in the convective envelope. As such, we believe that this extreme degree of pollution will have other effects on the stellar structure, which we discuss further in Section \ref{sec:discussion}.

\setlength{\tabcolsep}{5pt}
\begin{table*}
\caption{{\bf Mass in the CZ vs. Planetary Material.} The table lists the stellar mass in the outer CZ ($M_{CZ}$) for the considered range of stellar masses, based on the utilized {\tt MESA} models for the Main MS age (see Table \ref{tab:stellar_properties}), and compares this to the mass of the planetary material that is added to the CZ during the engulfment of Earth-like and Super-Earth planets ($m_p$ lost). Note that Earth-like planets are virtually entirely dissolved in the CZ for stars smaller than $1.3$~M$_\odot$. We also list the change in the measurable surface abundance of lithium, expressed in dex ($\Delta$~[Li/H]), from a singular engulfment event. A caveat for our results is that the added planetary mass is a significant fraction of the CZ mass, up to several times the CZ mass, for stars above a mass of about $1.4$~M$_\odot$, mainly due to the extremely small mass of the CZ for stars above the Kraft break. Additionally, this pollution is also added over an extended time of many hundreds of years (see Table~\ref{tab:engulf_times}). As such, we would not expect the CZ of these stars to maintain this extremely high degree of pollution for any significant extent of time, but rather exchange material with the underlying radiative layer, making our mass and abundance change estimates unrealistically high upper limits for these stars. The extremely large dex increases for [Li/H] (marked by asterisks) especially appear unrealistic. Another caveat is that we used the solar lithium values for our calculation of $\Delta$~[Li/H]; however, lithium is known to be deficient in the Sun, compared to nearby Sun-like stars \citep[e.g.,][]{Pasquini+1994}.}
\label{tab:mass_fractions}
\begin{flushleft}
\begin{tabular}{|c|c||c|c|c||c|c|c|}
\hline
    \multicolumn{2}{|c||}{} & 
    \multicolumn{3}{c||}{$1$~M$_{\oplus}$ Planet} & 
    \multicolumn{3}{c|}{$15$~M$_{\oplus}$ Planet} \\
\hline
$M_*$ (M$_{\odot}$) & 
$M_{cz}$ (g) & 
$m_{p}$ lost in CZ / $m_p$ &
($m_{p}$ lost) / $M_{cz}$ &
$\Delta$ [Li/H] &
$m_{p}$ lost in CZ / $m_p$ &
($m_{p}$ lost) / $M_{cz}$ &
$\Delta$ [Li/H] \\
\hline
0.5 & $3.09\times 10^{32}$ & 1 & $1.93\times 10^{-5}$ & 0.19 & 1 & $2.90\times10^{-4}$ & 0.95 \\
0.6 & $1.78\times 10^{32}$ & 1 & $3.36\times 10^{-5}$ & 0.28 & 1 & $5.04\times10^{-4}$ & 1.17 \\
0.7 & $1.32\times 10^{32}$ & 1 & $4.53\times 10^{-5}$ & 0.35 & 1 & $6.79\times10^{-4}$ & 1.29 \\
0.8 & $1.06\times 10^{32}$ & 1 & $5.64\times 10^{-5}$ & 0.41 & 1 & $8.46\times10^{-4}$ & 1.39 \\
0.9 & $7.27\times 10^{31}$ & 1 & $8.21\times 10^{-5}$ & 0.51 & 1 & $1.23\times10^{-3}$ & 1.54 \\
1.0 & $4.19\times 10^{31}$ & 1 & $1.42\times 10^{-4}$ & 0.69 & 1 & $2.10\times10^{-3}$ & 1.78 \\
1.1 & $1.58\times 10^{31}$ & 1 & $3.78\times 10^{-4}$ & 1.06 & 0.87 & $4.93\times10^{-3}$ & 2.20 \\
1.2 & $4.32\times 10^{30}$ & 1 & $1.38\times 10^{-3}$ & 1.59 & 0.33 & $6.76\times10^{-3}$ & 2.49 \\
1.3 & $1.04\times 10^{30}$ & 1 & $5.54\times 10^{-3}$ & 2.20 & 0.14 & $1.18\times10^{-2}$ & 2.87 \\
1.4 & $1.40\times 10^{29}$ & 0.38 & $1.62\times 10^{-2}$ & 2.87 & 0.0112 & $7.16\times10^{-3}$ & 3.44 \\ 
\hline
1.5 & $4.70\times 10^{26}$ & 0.0053 & $6.71\times 10^{-2}$ & 4.46$^{*}$ & 0.0106 & 2.02 & 5.91$^{*}$ \\
1.6 & $4.08\times 10^{25}$ & 0.0045 & $6.57\times 10^{-1}$ & 5.45$^{*}$ & 0.0095 & 20.7 & 6.95$^{*}$ \\
1.7 & $1.96\times 10^{25}$ & 0.0055 & 1.67 & 5.85$^{*}$ & 0.0117 & 53.8 & 7.30$^{*}$ \\
1.8 & $1.85\times 10^{25}$ & 0.0064 & 2.07 & 5.94$^{*}$ & 0.0138 & 67.0 & 7.33$^{*}$ \\
1.9 & $1.84\times 10^{25}$ & 0.0076 & 2.46 & 6.02$^{*}$ & 0.0164 & 80.1 & 7.34$^{*}$ \\
2.0 & $1.62\times 10^{25}$ & 0.0092 & 3.39 & 6.16$^{*}$ & 0.2009 & 111 & 7.41$^{*}$ \\
\hline
\end{tabular}
\end{flushleft}
\end{table*}

\subsection{Stellar Elemental Abundance Changes}

As the evaporated planetary mass is added to the stellar CZs, it alters the observable elemental composition of the stellar surfaces. We assume that the added material is mixed uniformly throughout the CZs and that the stars' initial compositions are solar. The $0.5$ to $1.3$~M$_\odot$ stars fully destroy the planet within their CZs. The $1.4$~M$_\odot$ star evaporates around 38\% of the planet in its CZ, thus only the planet's crust and around $60\%$ of its mantle are mixed into the CZ. The $1.5$ to $2$~M$_\odot$ stars only destroy less than $1\%$ of the planet in their CZs, meaning they only receive a fraction of the crust material. The masses of the polluting material are listed in Table \ref{tab:mass_fractions}. The amount of the crust, mantle, and core that the stars evaporate in their CZs determines which elements, and in what abundances, they mix into their CZs (see also Section \ref{subsec:planetcomp}). This determines the observability of the chemical enrichment of these engulfment events. 

In Table \ref{tab:dex_change1}, we show the number of singular Earth-like planets needed to produce a $0.1$~dex change for several key elements in the stars' surface composition after engulfment. We choose an abundance change of $0.1$~dex as a marginally detectable signal based on the general uncertainties in stellar abundance measurements observed with a variety of methods \citep[e.g.,][]{Hinkel+2016}.

We find that for stars smaller than $1.0$~M$_\odot$, tens to hundreds of Earth-like planets need to be engulfed in order to see this change. For stars between $1.0$ and $1.2$~M$_\odot$, one generally needs a few Earth-like planets, while for larger stars, less than a single Earth-like planet is required to produce a $0.1$~dex change. Notably, some elements are better suited to detect engulfment signatures than others. In particular, aluminium, calcium, and vanadium appear to be stronger markers of Earth-like planetary engulfment events, while oxygen, sodium, potassium, and manganese appear to be weaker markers; sulfur appears to be the weakest marker among those that we have tested. A special case here is lithium, for which we have listed the enrichment values from a single Earth-like planet engulfment in Table~\ref{tab:mass_fractions}. Lithium is often used as an indicator of stellar pollution due to the depletion of lithium in stars as a result of nuclear fusion processes; lithium enrichment is thus basically always measurable. However, our calculated lithium enrichment values may not be representative of the general stellar population, as the solar lithium abundance that we use is known to be deficient compared to close-by solar analogs \citep[e.g.,][]{Pasquini+1994}.

\definecolor{r}{HTML}{D55E00} 
\definecolor{y}{HTML}{F0E442} 
\definecolor{g}{HTML}{009E73} 
\definecolor{w}{rgb}{1.0,1.0,1.0}
\newcolumntype{C}{>{\columncolor{white}\centering\arraybackslash\hspace{5.25pt}}c<{\hspace{5.25pt}}}
\setlength{\tabcolsep}{0pt}

\begin{table*}[htbp]
\caption{{\bf Detectability of Stellar Elemental Abundance Changes due to Planetary Engulfment.} The table shows the number of Earth-like planets that need to be engulfed by a MS star of a given mass to produce a $0.1$~dex change in the observable surface concentration of a variety of elements common in rocky, terrestrial planets. The initial stellar composition was assumed to be solar. We do not list estimates for stars above $1.4$~M$_\odot$, as their surface CZs are exceedingly thin and the surface composition changes largely depend on the atomic diffusion and settling within the radiative layer of the star, and may thus be very short lived. The cell colors mark the pollution detectability; elements colored green are easiest to detect and only require a single Earth-like planet or less to be engulfed for a $0.1$~dex change; yellow marks intermediate detectability, requiring up to $15$~Earth-like planets \citep[see also][]{Oh+2018}; and orange marks low detectability, requiring more than $15$~Earth-like planets. Note that sulfur, iron, and nickel are less detectable in the $1.4$~M$_\odot$ case than in the $1.3$~M$_\odot$ case (see pale cells) since these elements come predominantly from the planet's core, which is not destroyed in the CZ for higher mass stars.}
\label{tab:dex_change1}
\begin{flushleft}
\begin{tabular}{|C|C|C|C|C|C|C|C|C|C|C|C|C|C|C|C|}
\hline
\vtop{\hbox{\strut $\, \, \, M_*$} \hbox{\strut (M$_{\odot}$)}} 
& O & Na & Mg & Al & Si & S & K & Ca & Ti & V & Cr & Mn & Fe & Ni & Y \\
\hline 
\rowcolor{r!70!w}
\cellcolor{w}0.5 & 252.8 & 188.6 & 56.40 & 42.17 & 54.16 & 692.2 & 117.5 & 43.48 & 48.13 & 37.64 & 47.52 & 174.8 & 51.57 & 53.21 & 45.37 \\
\rowcolor{r!70!w}
\cellcolor{w}0.6 & 145.7 & 108.7 & 32.51 & 24.31 & 31.22 & 398.9 & 67.70 & 25.06 & 27.74 & 21.69 & 27.39 & 100.7 & 29.72 & 30.67 & 26.15 \\
\rowcolor{r!70!w}
\cellcolor{w}0.7 & 108.1 & 80.61 & 24.11 & 18.02 & 23.15 & 295.8 & 50.20 & 18.58 & 20.57 & 16.09 & 20.31 & 74.69 & 22.04 & 22.74 & 19.39 \\
\rowcolor{r!70!w}
\cellcolor{w}0.8 & 86.78 & 64.75 & 19.36 & \cellcolor{y!60!w}14.48 & 18.59 & 237.6 & 40.32 & \cellcolor{y!60!w}14.93 & 16.52 & \cellcolor{y!60!w}12.92 & 16.31 & 59.99 & 17.70 & 18.27 & 15.58 \\
\rowcolor{y!60!w}
\cellcolor{w}0.9 & \cellcolor{r!70!w}59.6 & \cellcolor{r!70!w}44.47 & 13.3 & 9.94 & 12.77 & \cellcolor{r!70!w}163.2 & \cellcolor{r!70!w}27.69 & 10.25 & 11.35 & 8.88 & 11.21 & \cellcolor{r!70!w}41.20 & 12.16 & 12.55 & 10.70 \\
\rowcolor{y!60!w}
\cellcolor{w}1.0 & \cellcolor{r!70!w}34.35 & \cellcolor{r!70!w}25.63 & 7.67 & 5.73 & 7.36 & \cellcolor{r!70!w}94.03 & \cellcolor{r!70!w}15.96 & 5.91 & 6.54 & 5.12 & 6.46 & \cellcolor{r!70!w}23.75 & 7.01 & 7.23 & 6.17 \\
\rowcolor{y!60!w}
\cellcolor{w}1.1 & 12.95 & 9.66 & 2.89 & 2.16 & 2.78 & \cellcolor{r!70!w}35.43 & 6.02 & 2.23 & 2.47 & 1.93 & 2.44 & 8.85 & 2.64 & 2.73 & 2.33 \\
\rowcolor{g!80!w}
\cellcolor{w}1.2 & \cellcolor{y!60!w}3.54 & \cellcolor{y!60!w}2.64 & 0.79 & 0.59 & 0.76 & \cellcolor{y!60!w}9.68 & \cellcolor{y!60!w}1.65 & 0.61 & 0.68 & 0.53 & 0.67 & \cellcolor{y!60!w}2.45 & 0.73 & 0.75 & 0.64 \\
\rowcolor{g!80!w}
\cellcolor{w}1.3 & 0.85 & 0.64 & 0.19 & 0.15 & 0.19 & \cellcolor{y!60!w}2.33 & 0.4 & 0.15 & 0.17 & 0.13 & 0.16 & 0.59 & 0.18 & 0.18 & 0.16 \\
\rowcolor{g!80!w}
\cellcolor{w}1.4 & 0.20 & 0.14 & 0.05 & 0.03 & 0.05 & \cellcolor{r!40!w}\textbf{23.8} & 0.07 & 0.04 & 0.04 & 0.05 & 0.10 & 0.16 & \cellcolor{g!40!w}\textbf{0.30} & \cellcolor{g!40!w}\textbf{0.65} & 0.03 \\
\hline
\end{tabular}
\end{flushleft}
\end{table*}

\section{Discussion} \label{sec:discussion}

The results of our model calculations provide us with sensible estimates for the detectability of Earth-like planetary material pollution signatures for MS stars of a variety of masses. Most crucially, we have found that smaller stars with deep outer CZs are poor targets for observing chemical enrichment, often requiring hundreds of Earth-like planets to be engulfed to produce an observable signature. In contrast, stars in the $1$ to $1.4$~M$_\odot$ range appear to be suitable targets for finding such signatures, as generally, only a few Earth-like planets need to be engulfed to produce this signature.

\begin{figure*}
    \centering
    \includegraphics[width=1.0\linewidth]{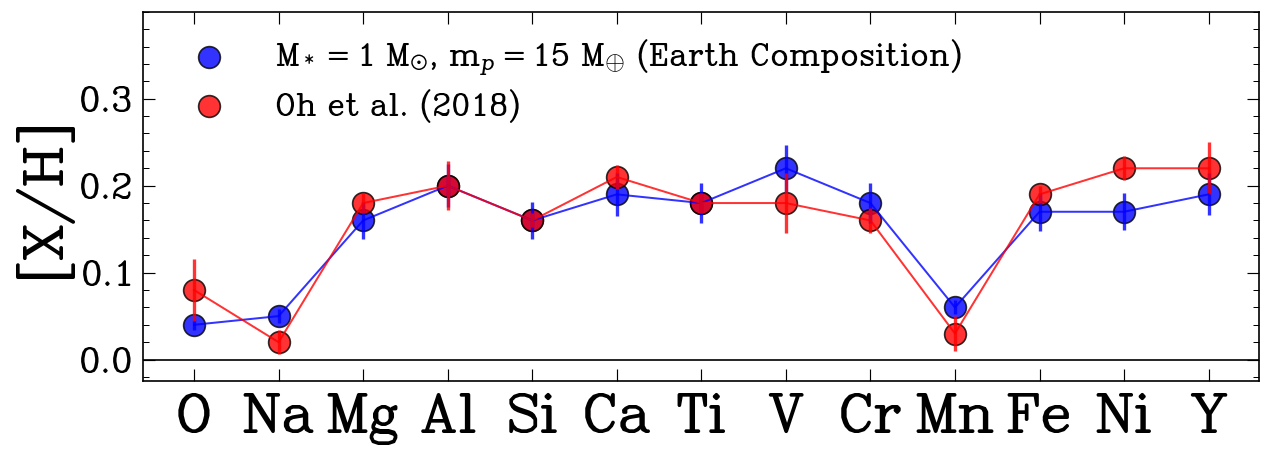}
    \includegraphics[width=0.5\linewidth]{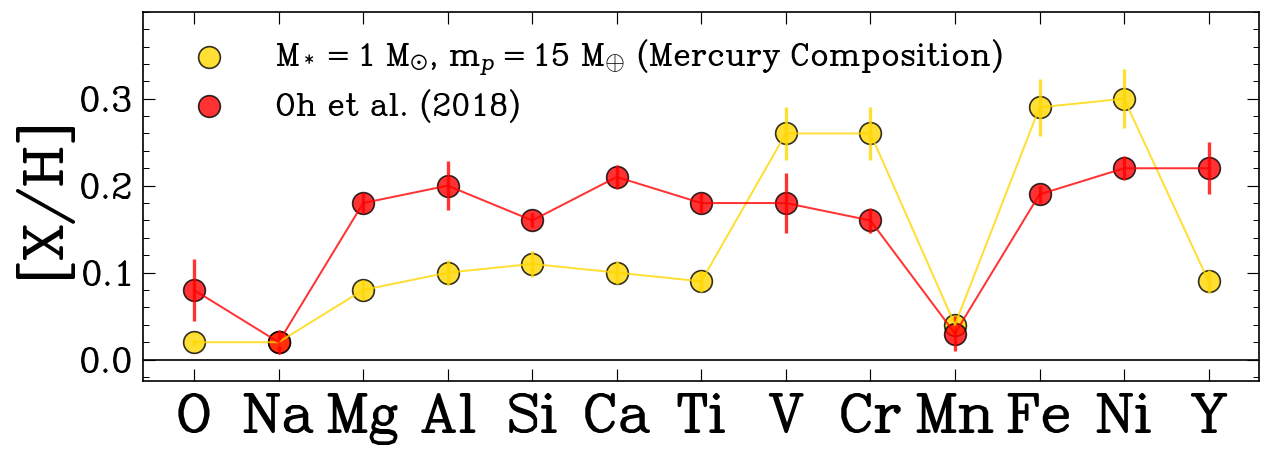}\includegraphics[width=0.5\linewidth]{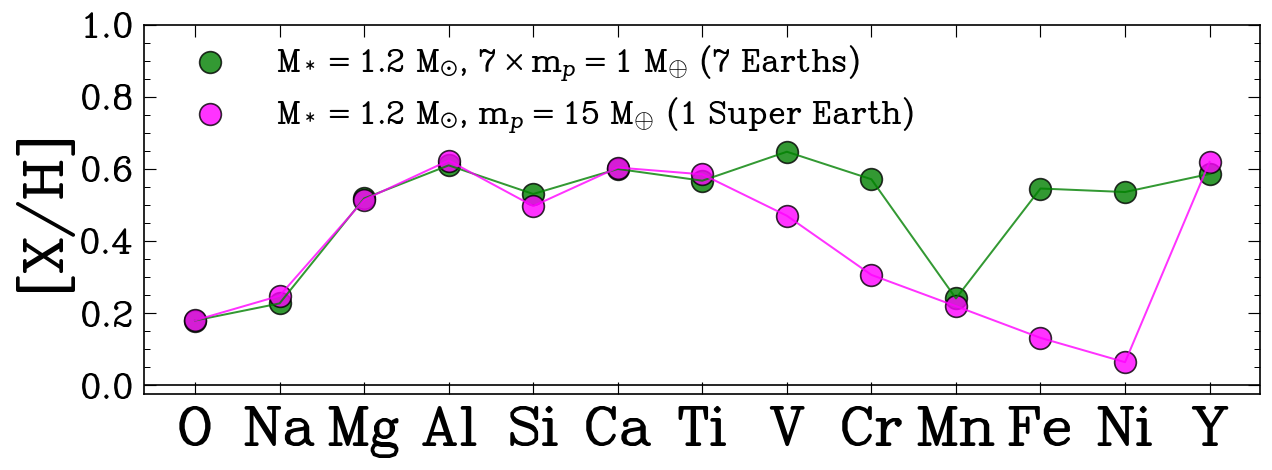}
    \caption{{\bf Comparison of our Model Results to Measured Abundance Enhancements for the Kronos-Krios Binary Pair.} The figures compare the measured [X/H] differences in elemental composition between the binary stars Kronos and Krios from the observations by \citet{Oh+2018} (red points) with the estimated elemental abundance enrichment of a star by planetary material. The large upper plot shows our model results for a $1$~M$_\odot$ solar-type star polluted by a $15$~M$_\oplus$ Super-Earth planet (blue points), which is consistent with the estimate by \citet{Oh+2018} that Kronos was polluted by $15$~M$_\oplus$ of planetary material with Earth-like composition. The lower left plot shows the pollution pattern produced by a $1$~M$_\odot$ solar-type star polluted by $15$~M$_\oplus$ of planetary material with a Mercury-like core mass fraction of about $\sim70~\%$ (gold points), which is much more weighted towards vanadium, chromium, iron, and nickel than the observed \citet{Oh+2018} values. The lower right plot shows the pollution patterns for a $1.2$~M$_\odot$ F-type star polluted by either a single $15$~M$_\oplus$ planet (magenta points), or $7$ separate $1$~M$_\oplus$ planets (green points), all with Earth composition, highlighting that for such stars, a larger planet will generally not add its core material to the star's CZ. The lines connecting the points only serve to guide the eye and have no meaning. Uncertainties for our simulated values were estimated using the range of CZ masses across the stars' MS life stage, but were ignored in the lower right plot for clarity.}
    \label{fig:dex_plot}
\end{figure*}

An important caveat here is that we assume an Earth-like planet composition and a solar-like stellar composition for our calculations. While this assumption can be justified based on the measured bulk composition of planetary material in WD pollution \citep[e.g.,][]{Bonsor2024}, we further support this choice by comparing our model results to the example case of the Kronos and Krios binary pair \citep[][]{Oh+2018}. The surface composition of Kronos, an approximately solar-type star, is measurably enriched in rocky planetary material compared to its companion Krios, with the polluting mass estimated to be about $15$~M$_\oplus$ of Earth-like planetary material. We calculate the enrichment signatures predicted by our model, assuming likewise a $15$~M$_\oplus$ planet with Earth-like composition, and compare our result to the observed enrichment pattern for Kronos in Fig.~\ref{fig:dex_plot}. Our result is, within uncertainties, consistent with the \citet{Oh+2018} measurements, supporting the validity of our approach. Furthermore, we determine that the pollution signature would be qualitatively different for an altered planet composition, via the example of a Mercury-type planet with a larger core mass fraction than Earth ($\sim 70~\%$ for Mercury vs. $\sim 30~\%$ for Earth), 
shown in the lower left plot in Fig.~\ref{fig:dex_plot}. We also note here that we cannot distinguish between the two possible engulfment scenarios of either one single Super-Earth with a mass of $15$~M$_\oplus$ or $15$ separate $1$~M$_\oplus$ planets for the \citet{Oh+2018} example, since in either case, the planet would be fully destroyed within the CZ of a Sun-like star like Kronos. For $1.2$ to $1.4$~M$_\odot$ stars, however, such a Super-Earth would only add its crust and part of its mantle to the CZs, thus potentially producing a different enrichment signature when compared to pollution by multiple small planets; the pollution would be less pronounced in vanadium, chromium, iron, nickel, and sulfur compared to the other rocky planet elements, as we show in the lower right plot in Fig.~\ref{fig:dex_plot} (see also the $1.4$~M$_\odot$ case in Table~\ref{tab:abundances}).

For stars larger than about $1.4$~M$_\odot$, while the polluting material is a significant, often massive, fraction of the CZ mass, we caution that such pollution signatures may not be as straightforward to observe. For one, these stars, being mostly radiative, do not have significant CZs, and the churning up of their surface from centuries of direct interaction with the engulfed planet may have a non-negligible effect on the stellar surface structure. Additionally, since the CZs of these stars are very thin, processes such as atomic diffusion, convective undershoot, or potentially thermohaline mixing may deplete any planetary material added to the surface by delivering it to the underlying radiative layer over time. We note here that the rate of planetary material added to the surface equals roughly $10^{-3}$ times the mass of the outer CZ per year, over a span of $\approx 500-1000$ years; as such, relatively little mixing or settling would be required to deplete this material before it could build up significantly. 

Another important caveat is that our model assumes an initially circular planetary orbit right beneath the stellar surface, and we do not consider what kind of mechanism may have brought the planet there. As such, our calculated engulfment durations may have significant uncertainties, since various migration mechanisms may have produced different initial conditions than we have considered here. For example, if the planet migrated due to tides, the star would have had an altered rotation rate, most likely reducing the relative velocity between the planet and the stellar envelope, which could extend the engulfment process duration. If, instead, the planet entered the stellar envelope on an eccentric orbit, the relative velocity between the planet and stellar envelope at periastron would most likely be larger, increasing the experienced drag and potentially reducing the engulfment duration. Furthermore, the added planetary material and interactions may alter the envelope's scale height and drag characteristics.

Given the mentioned caveats, future work on planetary engulfment will require a more careful treatment of several of the processes modeled here. For example, the additions to the envelope of planetary material, angular momentum, and energy from the drag interaction may all alter the stellar structure of the star significantly enough to drive, for example, internal mixing processes and stellar structure changes. It has already been observed that planetary engulfment events may cause the formation of a cool, dusty ejecta shroud around the star, as well as an accretion disk from material falling back onto it \citep{Lau+2025}. We plan to more accurately model these changes in the future via a suite of {\tt MESA} models. Additionally, we have so far modeled the planet with uniform density, which isn't realistic given the radial composition and density profiles of rocky planets, and should be replaced with a polytropic or multi-layered planetary model. We have also ignored the potential contribution to the planet's evaporation and destruction via the latent heating from the stellar radiation. This approach may be suitable for the relatively rapid radial migration of the planets in the smaller mass stars, but may be inadequate for the larger stars, where the engulfment process can take hundreds to thousands of years and radiative heating of the planet and its interior should be considered. Finally, a more careful treatment of the planet's orbit and migration history before entering the star is necessary to determine more accurate estimates of the engulfment process duration.

\section{Summary and Conclusion}\label{sec:conclusion}

In this work we have presented an analytical model of the engulfment of Earth-like planets by MS stars in the mass range of $0.5$ to $2.0$~M$_\odot$ in order to gauge the observability of the resulting chemical pollution signatures. Our model tracks the planetary orbital evolution, evaporation, and destruction, and considers the stellar and planetary structures and compositions. We have tested our model results against some previous observations of planetary engulfment chemical signatures, in particular the case of the Kronos-Krios binary, and found them to be consistent (see Fig.~\ref{fig:dex_plot}). Overall, our model results have yielded several key insights regarding engulfment signature observability:

\begin{enumerate}
    \item Stars in the $1.0$ to $1.4$~M$_\odot$ range are good candidates for measuring chemical enrichment signatures, as the CZs of these stars are of limited size and mass, and terrestrial planets will generally be completely destroyed within the CZs. Smaller stars have CZs that are very deep and massive, strongly diluting any pollution signatures. 
    \item Well-suited elements for observing rocky planet pollution signatures in MS stars include aluminium, calcium, vanadium, as well as lithium (which is a strong pollution indicator for planetary engulfment in general), while elements such as sulfur, oxygen, sodium, potassium, or manganese are ill-suited.
    \item For stars in the $1.2$ to $1.4$~M$_\odot$ range, it may be possible to determine whether a potential polluting body was a single large planet like a Super-Earth or several smaller Earth-like bodies. As the core of a larger planet may not be disrupted within the CZ, but can migrate below it before destruction, these different engulfment scenarios would produce different pollution signatures, in particular in vanadium, chromium, iron, nickel, and sulfur, compared to the other elements.
    \item Stars more massive than $1.4$~M$_\odot$ are above the Kraft break and have negligible CZs. Planetary engulfment events may significantly alter their stellar surface structure, making observational signatures uncertain. More work is required to understand these cases.
    \item The timescale of the engulfment process varies strongly with stellar and planetary mass, with durations of hundreds or thousands of years for more massive stars and Earth-like planets, and durations of only a few years for small stars and Earth-like planets. The engulfment timescale is generally shorter for Super-Earths, by about a factor of 8.
\end{enumerate}

Further work will need to be done to accurately explore the stellar surface composition changes for stars above the Kraft break, in particular the potential stellar structure changes, which is beyond the scope of this work.

Overall, the results of our model can serve as a guide for selecting suitable observational targets for future searches of planetary engulfment sites, and can help to gauge whether composition differences between binary stars or co-moving pairs can originate from planetary engulfment to begin with; i.e., for small stars with deep convective envelopes, it is unlikely that any measured chemical differences originate from planet engulfment, in the absence of other engulfment indicators such as increased stellar spin rates.

\section{Acknowledgments}
We thank the anonymous referee for their helpful comments and suggestions. K.T.L. is grateful for support from the Vanderbilt Initiative in Data-intensive Astrophysics (VIDA). A.P.S. acknowledges support from NASA grant 80NSSC24M0022. R.Y. is grateful for support from a Doctoral Fellowship from the University of California Institute for Mexico and the United States (UCMEXUS), a NASA FINESST award, and a Chancellor's Dissertation-Year Fellowship from UC Santa Cruz. The authors thank Girish Duvvuri, Smadar Naoz, Morgan MacLeod, and Ruth Murray-Clay for helpful comments and discussions.

\software{Matplotlib \citep{Hunter2007}, NumPy \citep{Harris+2020}, SciPy \citep{Virtanen+2020}, MESA \citep{Paxton+2011,Paxton+2013,Paxton+2015,Paxton+2018,Paxton+2019,Jermyn+2023}}

\bibliography{Lane_references}{}
\bibliographystyle{aasjournal}

\end{document}